# Investigation of the Status of Unit 2 Nuclear Reactor of the Fukushima Daiichi by the Cosmic Muon Radiography

May 12, 2020

- Hirofumi Fujii[1], Kazuhiko Hara[2], Shugo Hashimoto[2,%], Kohei Hayashi[1], Hidekazu Kakuno[3], Hideyo Kodama[1], Gi Meiki[6], Masato Mizokami[5], Shinya Mizokami[5], Kanetada Nagamine[1], Kotaro Sato[1], Shunsuke Sekita[5], Hiroshi Shirai[6], Shin-Hong Kim[2], Takayuki Sumiyoshi[3], Atsuto Suzuki[1,$], Yoshihisa Takada[2], Kazuki Takahashi[2,#], Yu Takahashi[2,&], Fumihiko Takasaki[1], Daichi Yamada[5] and Satoru Yamashita[4]

[1] *High Energy Accelerator Research Organization (KEK), Tsukuba, Ibaraki 305-0801, Japan*
[2] *University of Tsukuba, Tsukuba. Ibaraki 305-8571, Japan*
[3] *Tokyo Metropolitan University, Japan*
[4] *University of Tokyo, Japan*
[5] *Tokyo Electric Power Company Holdings, Incorporated[*
[6] *TEPCO systems corporation*

[%] Present address: JAXA
[$] Present address: Iwate Prefectural University
[#] Present address: Toshiba Co., Ltd.
[&] Present address: NEDO

## Abstract

We have investigated the status of the nuclear debris in the Unit-2 Nuclear Reactor of the Fukushima Daiichi Nuclear Power plant by the method called Cosmic Muon Radiography. In this measurement, the muon detector was placed outside of the reactor building as was the case of the measurement for the Unit-1 Reactor. Compared to the previous measurements, the detector was down-sized, which made us possible to locate it closer to the reactor and to investigate especially the lower part of the fuel loading zone. We identified the inner structures of the reactor such as the containment vessel, pressure vessel and other objects through the thick concrete wall of the reactor building. Furthermore, the observation showed existence of heavy material at the bottom of the pressure vessel, which can be interpreted as the debris of melted nuclear fuel dropped from the loading zone.

Subject Index: Fukushima-Daiichi; muon radiography; nuclear debris

## 1. Introduction

Following the investigation [1] made for the Unit-1 reactor of the Fukushima-Daiichi, the same team tried to investigate the Unit-2 reactor of the Fukushima-Daiichi by the same technique of the cosmic muon radiography [2], [3]. Although we have successfully demonstrated that the structure of the Nuclear Power Reactor can be visualized by the telescope placed outside of the reactor, we noticed the obtained images are less clear, especially at the lower part of the reactor. To obtain a close-up view of the area, we constructed a new telescope system and located it as close as possible to the building of the Unit-2 reactor. The new system is much smaller than the previous telescope, smaller than 1-m cubic, so that it enabled us to find an appropriate place for the muon radiography, namely to place it avoiding massive obstacles along the view line of objects to be investigated and ensuring necessary elevation angle to the objects.

Since the ambient radiation level around the nuclear reactors at the Fukushima-Daiichi has been significantly reduced thanks to the efforts by the TEPCO, radiation protection as much as was employed for the system for the Unit-1 reactor investigation (10-cmthick iron) has become not necessary. Having a thinner radiation protection plate made the telescope weight lighter, which helped to ease the installation work allowing us to place the telescope right next to the building wall.

## 2. The Muon Telescope

In the telescope employed for the Unit-1 reactor, we used 1-m long plastic scintillator bars with the cross section of 1 cm×1 cm [4], [5], [6]. They are arranged in a hodoscope covering an effective area of 1 m$^2$. For the new telescope, we used the same scintillator bars of 50-cm length instead and they covered an area of 50 cm×50 cm, one quarter of the previous telescope's area. We measured the hit positions of particles on the hodoscope plane in two orthogonal coordinates, X and Y coordinates. The telescope was composed of a pair of XY hodoscopes. The scintillator bars used for the telescope are shown in Fig. 1. Through the hole at the center, a plastic wavelength shifting fiber was inserted, which collected the scintillation light produced upon the passage of particles through the scintillator and sent it to a photo-sensor (Multi-Pixel Photon Counter, MPPC [7]) attached to the fiber end.

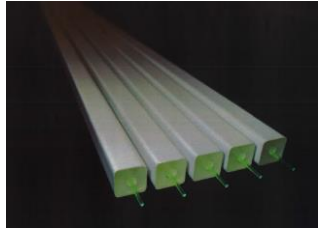

Fig. 1 Plastic scintillation bars of 1 cm$^2$ cross section with wavelength shifting fibers inserted through the holes.

Although the scintillator hodoscope area was reduced, the acceptance in terms of viewing angle was maintained since the distance between the XY hodoscopes was also reduced from 1.5 m of the Unit-1 detector to 0.7 m. In order to have a similar angle resolution of the muon flight direction under the shortened hodoscope distance, the position resolution was required to be halved while using the same scintillator bars. For this requirement, we took an approach by doubling the hodoscope layers with neighboring layers displaced by 0.5 cm, one half of the scintillator bar width, to each other as shown in Fig. 2. The configuration was intended to require coincidence between the two adjacent layers and hence the area traversed by the cosmic particles can be effectively halved. Therefore, the telescope has in total eight layers of scintillator planes to measure the trajectory of the cosmic muons.

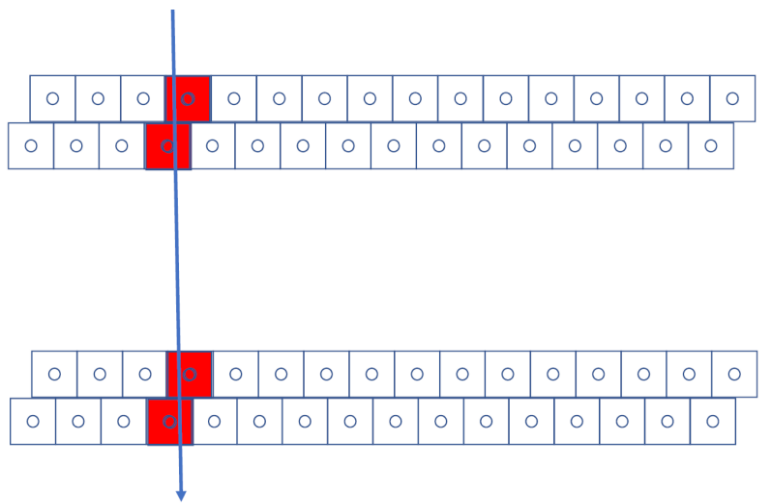

Fig. 2. Cosmic muon tracking in the telescope. Single hodoscope layer is composed of two layers of 1 cm wide scintillator bars staggered by 0.5 cm. The track direction is measured by two sets of hodoscopes separated by 70 cm. The illustration is referred to one coordinate only with another set running at 90° with respect to this coordinate (not shown). There are fifty 1-cm wide bars arranged in each layer. The circles at the center are for the wavelength shifting fibers.

Figure 3 shows a photograph of one set of XY hodoscope planes. The blue cables running out in two directions are used to read out the signals of 200 MPPCs to FPGA-based readout systems [1], [4], [5]. The grey cables are used to feed bias voltages to MPPCs with the bias voltages adjusted automatically to maintain the same MPPC gains accounting for the temperature variation [7].

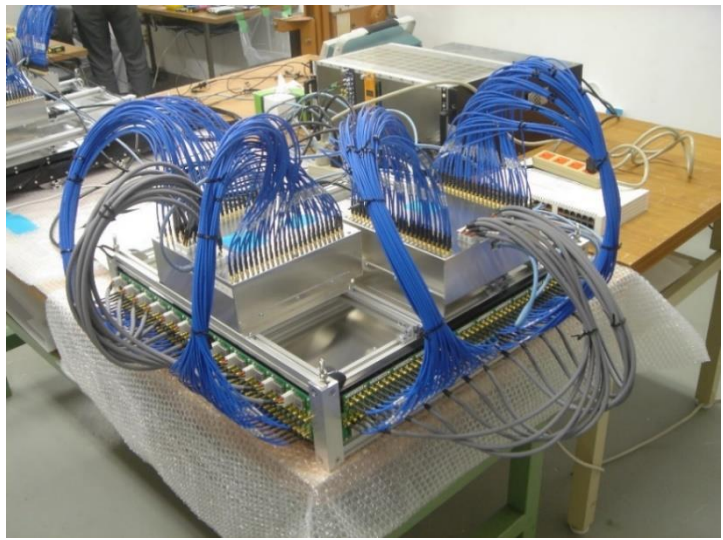

Fig. 3. One of the assembled XY hodoscope units with data acquisition electronics mounted.

The performance of the XY hodoscope as a telescope was first tested at KEK. The two assembled XY hodoscope units were set horizontally separated by 50 cm in vertical direction. An iron block of 20 cm cubic was placed above them to examine the sensitivity of the cosmic muons.
Figure 4 shows the observed rate of cosmic muons plotted as a function of the hit position in the two orthogonal coordinates. The dips in the rates are due to the iron block of 20 cm thickness. The distributions corresponding to the block edges are used to estimate the position resolution, which resulted as good as 0.5 cm, as expected.

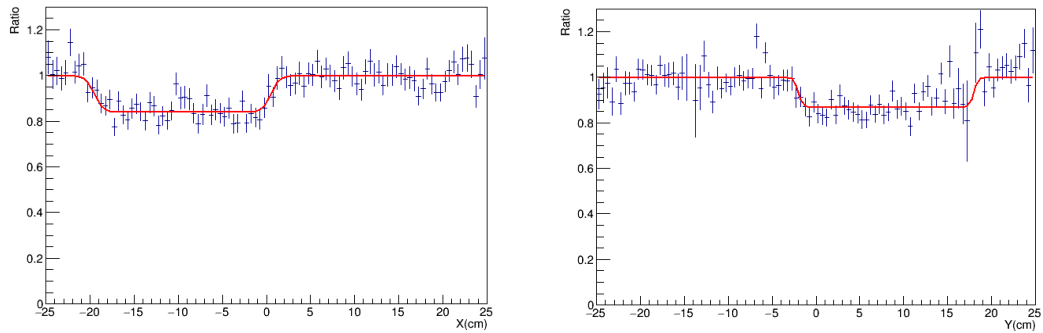

Fig. 4. Cosmic muon rates (in Hz) observed with two telescope systems shown in Fig. 3.

The two XY hodoscope units were assembled with the front unit elevated by 10 cm to the rear unit such that the center of the viewing angle is targeted at the center of the Reactor Pressure Vessel (RPV). The assembled telescope was housed in a container made of 2-mm thick aluminum plates as shown in Fig. 5. The size of the aluminum container was 85 cm×108 cm×125 cm. The total weight was about 270 kg. A schematic image of the scintillation counter arrangement in the telescope is given in Fig.6.

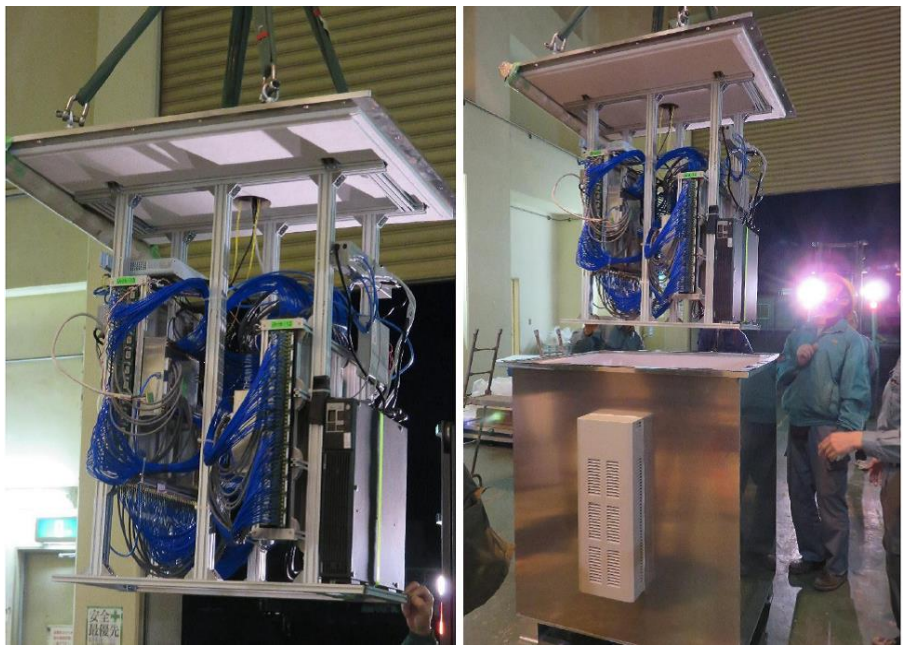

Fig. 5. The assembly of the telescope to be placed in the aluminum container. The container is equipped with an air-conditioner (the air exchange box is seen).

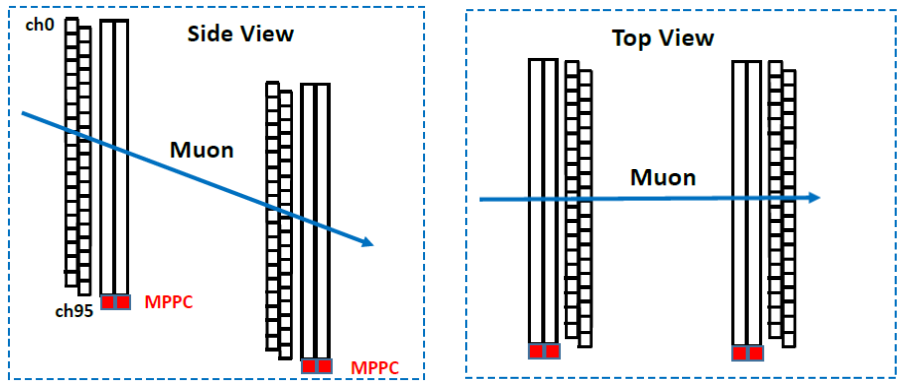

Fig. 6. Counter arrangement of the muon telescope, the side view and top view. The actual number of scintillator bars is fifty per plane. The scintillator signals are read out by MPPCs attached to one end of the bar.

## 3. Investigation Results of the Unit-2 reactor

*3.1 Telescope installation*

The telescope was placed at the Fukushima Daiichi facing right to the Unit-2 reactor building where the radiation level was as low as 100 μSv/h. The ambient radiation level had been significantly reduced since the Unit-1 measurement. The telescope container made of 2 mm aluminum was covered by two 2-mm thick lead sheets further to reduce the in-box radiation level to 20 μSv/h. The lower ambient radiation level together with the requirement of eight-layer coincidence in tracking made us possible to use such thin radiation shield to the detector.

Figure 7 shows the location of the muon telescope viewing the Unit 2 reactor. The red arrow is the direction of the telescope's view center targeting at the center of the reactor.

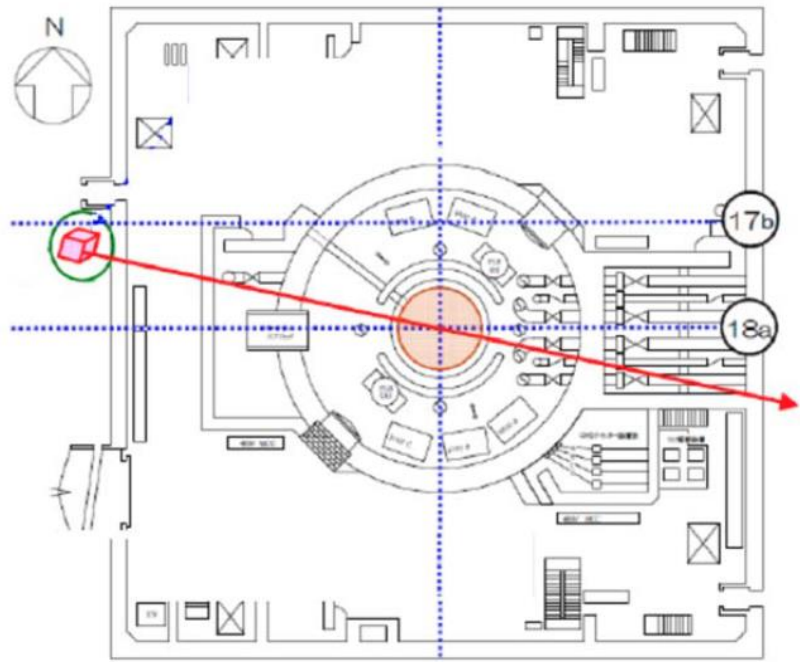

Fig. 7. Top-view showing the location of the telescope at the Unit-2 reactor building. The telescope system is shown in the red box circled in green.

*3.2 Images of the Unit-2 Reactor*

We started the measurement in March 17, 2016 and continued till September 17, 2016. The number of events observed by the telescope was about 7 million. The obtained image of the Unit 2 reactor is shown in Fig. 8. The figure shows the amount of material traversed by the cosmic muons in the unit of density-length (g/cc-m). The procedure of deriving the amount in density-length is described in [1].

The wall of the Primary Containment Vessel (PCV), which is made of at least 1.7 m thick heavy concrete of density of about 2.3 g/cm$^3$, is characteristic and imaged successfully. Compared to the PCV, the wall of the RPV made of approximately 14 cm iron is much thinner both in physical and density lengths, especially the material thickness projected along the muon path. Therefore, the observation does not show a distinct image of the RPV, compared to the calculated distribution where muon scattering is not included.

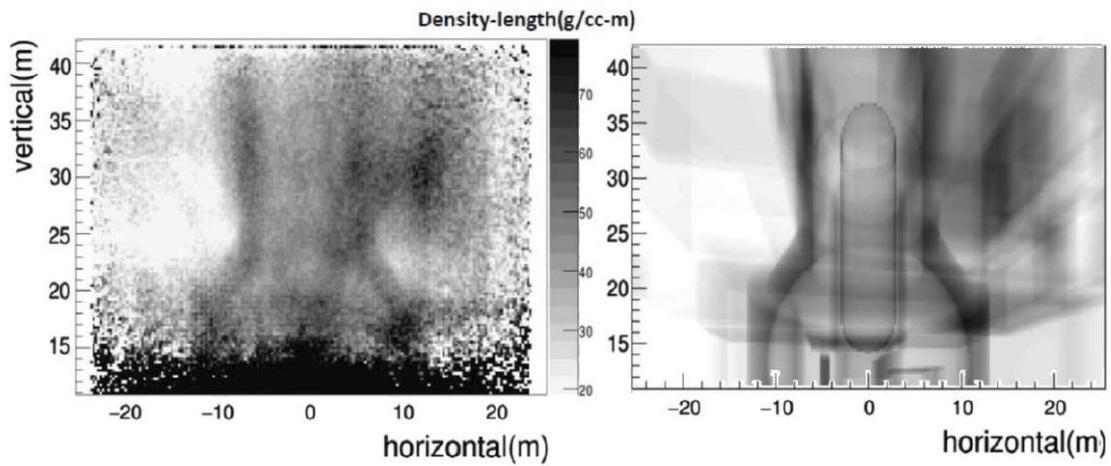

Fig. 8. (Left) An image of density-length distribution of the Unit-2 reactor obtained by this measurement and (right) an image calculated (no muon scattering included) under assumption the nuclear fuel melt and dropped from the loading zone.

Figure 9 illustrates the zoomed image around the RPV in comparison with its drawing. The pixel size of the image corresponds to 25 cm× 25 cm at the distance of the loading zone center. We note that any sign of spots of heavy attenuation is not identified in the center of the RPV (the area centered horizontally and 19-23 m vertically) where the nuclear fuel was loaded originally. The observation can conclude that most of the nuclear fuel had melt and dropped to the lower level. Actually, there exits an area where cosmic rays are attenuated strongly as indicated by the dark spots below 16 m in the vertical direction.

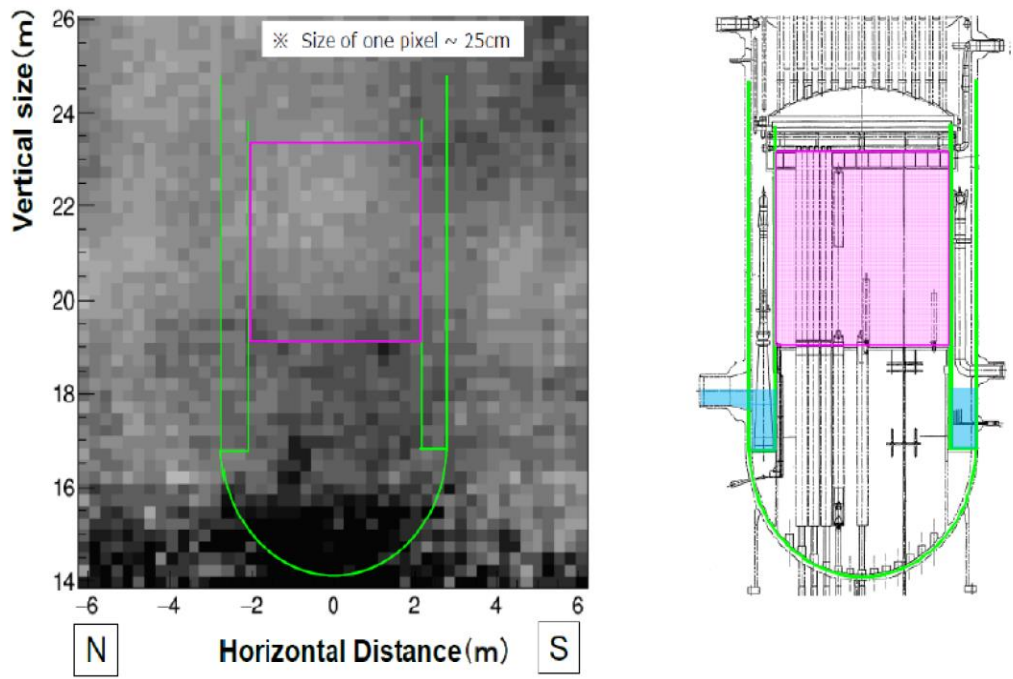

Fig.9. A zoomed image around the RPV in comparison with its drawing (the drawing provided by the TEPCO). The RPV wall is shown in green. The region shown in pink is the fuel loading zone. The vertically running structures under the loading zone shown in the drawing are the reactor control rods.

*3.3 Estimation of the amount of the nuclear debris remaining in the RPV*

The observed strong attenuation of the cosmic muons could be attributed to the debris of the melted nuclear fuel. We estimated the weight of the remaining substance in two methods. Method-1 calculates the weights outside the RPV based on the drawings of the plant and subtracts their contribution from the measured absorption. Method-2 is the procedure to evaluate the weights by subtraction of the side-band absorption, the procedure essentially identical to the Unit-1 investigation [1].

*3.3.1 Method-1 Estimation*

The horizontal density-length distributions were calculated from the plant drawing in the four height slices as illustrated in Fig 10. The calculation results are shown in Fig. 11 where measured density-length values with subtracting the amount outside the RPV are overlaid. In the calculation, the average density inside the RPV was assumed to be either zero (nothing is remaining), 2 g/cc, or 6 g/cc.

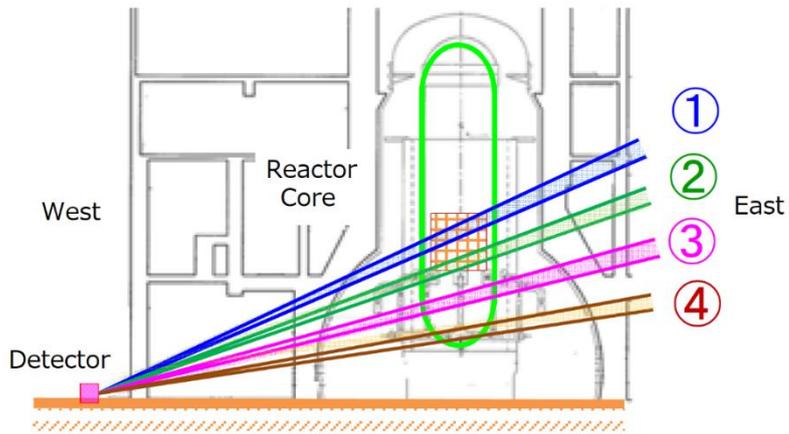

Fig.10. Vertical slices to evaluated the density-length distributions used in Method-1. Slice-1 and Slice-2 are for the upper and lower parts of the fuel loading zone, Slice-3 is below the loading zone, and Slice-4 is for the bottom of the RPV. The slice height is 50 cm at the distance at the center of the RPV.

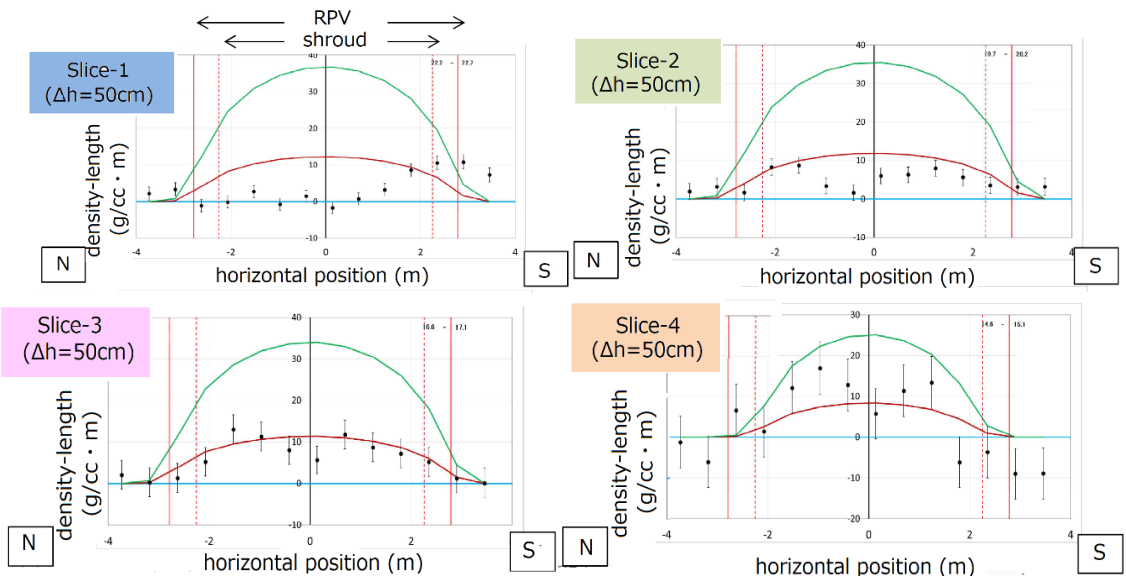

Fig.11. The density-length distributions with subtracting the materials outside the RPV are plotted for the four vertical slices defined in Fig. 10. In the calculations, the average density inside RPV was assumed to be zero (blue), 2 g/cc (red), or 6 g/cc (green). The density-length values estimated from the measured data are plotted in black points. The vertical solid (dotted) lines at ±2.5 m (±2.0 m) indicate the boundary of the RPV (reactor core shroud).

In Slice-1 (upper part of the fuel loading zone), the vast region is consistent that there remains nothing. In Slice-2 and Slice-3, the average density is 1 to 2 g/cc, and substantial material appears to remain inside the RPV. In Slice-4, there seems to remain more dense material, although the statistical significance is weaker. We also note that the agreement in the regions outside the investigation is fairly good, demonstrating the reliability of the present method.

The same technique was used to evaluate the weight in the RPV, separating the entire RPV into three regions; A) volume above the loading zone, B) volume of the loading zone, and C) volume below the loading zone. The estimation results are summarized in Table 1 in comparison with the weights calculated for the undamaged case.

The uncertainty is mostly from the limited knowledge of the muon momentum distribution and uncertainty in the material calculation along the muon path. The breakdown of the uncertainty is explained for Method-2.

TABLE 1. The amount of material inside the RPV estimated by Method-1, in comparison with the values calculated for the undamaged case. The uncertainty numbers are for statistical and systematic ones. The estimations based on Method-2 are also listed.

| Volume in RPV | Weight (tons) Method - 1 | Weight (tons) Method – 2 | Undamaged Assumption weight (tons) |
|---|---|---|---|
| above the loading zone: A) | 96±6±29 | 134±5±35 | 80 for Method-1<br>114 for Mothod-2 |
| in the loading zone (within shroud) : B) | 49±5±33 | 17±3±19 | 160 (fuel assembly)<br>15 (control rods) |
| Below the loading zone: C) | 159±9±51 | 197±10±60*<br>156±8±57 | 35<br>excluding water |
| B) + C) | 208±11±60 | 173±9±60 | 210 |

3.3.3 *Method-2 Estimation*

We divide the image into several regions. The region in |x|<2 m is named "a" where we investigate the amount inside the RPV, and the region between x=-5 m and x=-2 m named "b" is the sideband used to evaluate the contributions outside the RPV. This method is identical as used in the investigation for the Unit-1 reactor [1], and the right sideband was not used as the influence of the fuel storage pool was significant. The two horizontal regions are further divided vertically each into three according to the height. Note the definition of the volume A) was not identical to Method-1, therefore different calculated values are given in the last column. Table 1 summarizes the estimated amounts in tons for the difference between the areas "a" and "b". Here we employ "difference" in evaluation as the main systematics due to muon momentum spectrum uncertainty [1] is reduced significantly in the

difference. The quoted systematic uncertainties are explained in the following section. The central value for Volume C) is corrected to 156 tons, which is also explained in the next section.

As the material in the volume A) above the loading zone is considered unchanged by the accident, the measured amounts and the amounts calculated from the available drawing information can be used to evaluate the reliability of the present evaluations. The calculated "a"-"b" as 114±6(sys) tons is consistent with the measurement.

*3.3.4 2 Systematic uncertainty*

In order to verify the uncertainty of the present measurement especially in the lower height area, we tried to measure the amount of materials of a known object. For this purpose, we chose the lower part of the PCV concrete wall. The two areas of the PCV as marked by the red square in Fig. 10 are defined within the height interval between 14.3<h<19.5 m and horizontal interval between 8.9 <|x|< 13.3 m about the PCV center. The material other than the RVC walls was estimated in the areas of the same horizontal level but outer adjacent to the red squares (13.3 <|x|< 16.5 m, pink dotted regions).

The measured amounts of the PCV wall are 375±11 tons and 362±11 tons for the left-side and right-side walls, respectively. The average 369 tons should be compared to the weight 292 tons calculated from the drawing geometry for a concrete density of 2.3 g/cc. As the difference is probably due to insufficient knowledge of the muon momentum distribution in this zenith angle range, we take these numbers to derive the correction factor 292/369 = 0.79. This factor is used to correct for the value measured in the bottom of the fuel loading zone C), see Table 1. The relative difference of the correction factor is added in quadrature to the other uncertainties.

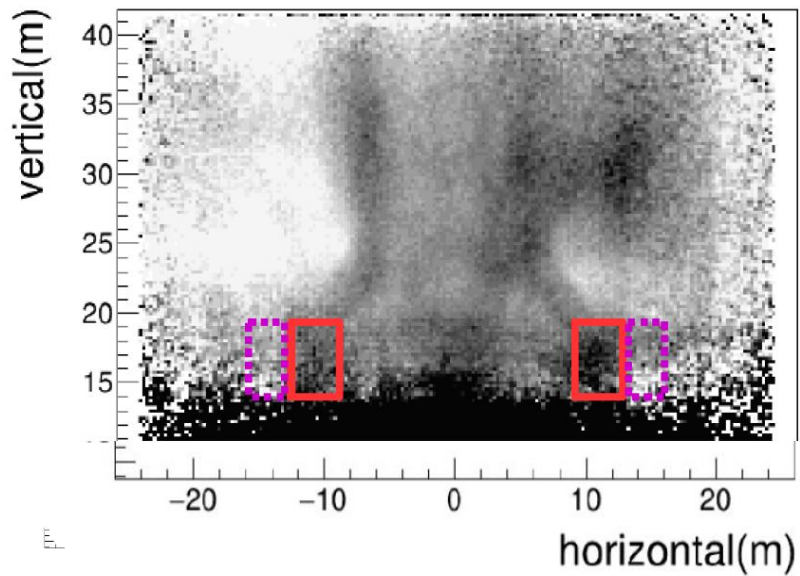

Fig.10. (Red solid rectangles) Areas to evaluate the amount of material in the bottom part of the RPV. (Pink broken rectangles) Adjacent areas to estimate the material other than the RPV structure.

As shown in Table 1, the measurement and calculation values are consistent for the upper parts of the loading zone. The difference is taken also as a systematics. All the contributions to the systematic uncertainties are summarized in Table 2. Among these, muon momentum modeling is related to the limited knowledge of the muon flux. The detector threshold concerns the difference of the detector efficiencies measured at KEK and Fukushima-Daiichi. The "a"-"b" uncertainty is as explained above for the difference between the measurements and calculations in the regions above the loading zone. The region definition concerns about the definition of the "a" and "b" regions, where the difference when the regions were defined one bin (=25 cm) offset horizontally or vertically.

TABLE 2. Contributions to the systematic uncertainties. Units are in ton.

| Contribution | 14.3<h<19.5m Bottom of L.Z. | 19.5<h<23.5m Loading zone | 23.5<h<28.9m Above the L.Z | 23.5<h<31.0m Above the L.Z |
|---|---|---|---|---|
| Muon momentum spectrum modeling | 20 | 3.4 | 17.8 | 22.3 |
| Detector threshold | 28 | 9.5 | 7.1 | 10.0 |
| "a"-"b" uncertainty (Table 1) | 41 | 4.4 | - | - |
| Region definition | 21 | 13.0 | 14.5 | 17.5 |
| Calculation from drawing | 2 | 9.0 | - | - |
| Total | 57 | 19.3 | 23.5 | 29.1 |

## 4. Summary

We investigated the status of Unit-2 Nuclear Reactor of the Fukushima-Daiichi by the cosmic muon detectors placed outside of the reactor building. The detector was down-sized from the system deployed for the Unit-1 observation to almost 1 m cubic. The obtained image is consistent that most of the nuclear fuel assemblies do not exist in the original location. We evaluated the amount of the materials left in the fuel loading zone is 17 – 49 tons. The amount found in the lower part of the pressure vessel is about 160 tons. The amount of the fuel assemblies originally at the loading zone is estimated to be 160 tons, therefore the observation is consistent that the most of the fuel debris are located at the bottom of the pressure vessel. We have demonstrated that the cosmic muon radiography is very effective to locate the heavy object inside a large complex object and to measure exclusively the amount of material in weight.

## Appendix Weight distribution in the RPV

The weight distirbution inside the RPV was derived using Method-1 by diving the area into projective towers with 50 × 50 cm cross sction at the center of the RPV. The result is given in Fig. A where heavier projections are shown brighter. Also shown is a possible status of the Unit-2 PCV guessd using the derived distribution. The red clusters represent the debris.

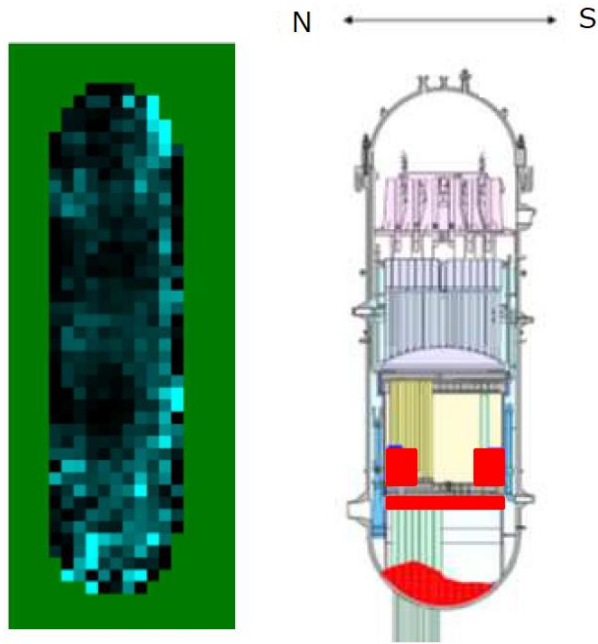

Fig. A. (left) Mass distribution inside the RPV. (right) A drawing showing possible status of the Unit-2 RPV status. The clusters in red represent the nuclear debris.

## Acknowledgements

The present study has been carried out as a project supported by the International Research Institute for Nuclear Decommissioning IRID and by the TEPCO. The cooperation and lots of help provided by the TEPCO and Tokyo Power Technology Ltd. have been inevitable for the detector installation and operatoin. We thank Professor Hirotaka Sugawara of KEK for many advices and suggestions.